\documentclass[a4paper]{jpconf}

\usepackage{graphicx}
\usepackage{subfigure}
\usepackage{amsmath,color}
\def\gappeq{\mathrel{ \rlap{\raise.5ex\hbox{$>$}}
                      {\lower.5ex\hbox{$\sim$}} } }
\def\lappeq{\mathrel{ \rlap{\raise.5ex\hbox{$<$}}
                      {\lower.5ex\hbox{$\sim$}} } }

\usepackage{color}

\newcommand{\del}[1]{\textcolor{red}{}}

\begin{document}

\title{Vortices in the two-dimensional dipolar Bose gas}

\author{B. C. Mulkerin$^1$, D. H. J. O'Dell$^2$, A. M. Martin$^1$ and N. G. Parker$^3$} 
\address{$^1$ School of Physics, University of Melbourne, Victoria 3010, Australia

$^2$ Department of Physics and Astronomy, McMaster University, Hamilton, Ontario, L8S 4M1, Canada

$^3$ Joint Quantum Centre Durham--Newcastle, School of Mathematics and Statistics, Newcastle University, Newcastle upon Tyne, NE1 7RU, United Kingdom}
\ead{nick.parker@newcastle.ac.uk}

\begin{abstract}
We present vortex solutions for the homogeneous two-dimensional Bose-Einstein condensate featuring dipolar atomic interactions, mapped out as a function of the dipolar interaction strength (relative to the contact interactions) and polarization direction.  Stable vortex solutions arise in the regimes where the fully homogeneous system is stable to the phonon or roton instabilities.  Close to these instabilities, the vortex profile differs significantly from that of a vortex in a nondipolar quantum gas, developing, for example, density ripples and an anisotropic core.  Meanwhile, the vortex itself generates a mesoscopic dipolar potential which, at distance, scales as $1/\rho^2$ and has an angular dependence which mimics the microscopic dipolar interaction.
\end{abstract}

\section{Introduction}
A landmark experiment in quantum gas research was the achievement in 2005 of a dipolar Bose-Einstein condensate (BEC) made of $^{52}$Cr atoms \cite{Griesmaier}.  Other groups have subsequently used novel methods to make $^{52}$Cr  condensates \cite{Beaufils}, as well as dipolar condensates composed of $^{164}$Dy \cite{Lu} and $^{168}$Er \cite{Aikawa}.  The dipolar nature of these atoms arises from their sizeable magnetic dipole moment which for $^{52}$Cr is $d=6 \mu_{B}$, for $^{168}$Er is $d=7 \mu_{B}$, and for $^{164}$Dy is $d=10 \mu_{B}$,  where $\mu_{B}$ denotes the Bohr magneton. These numbers should be compared to the value $d=1 \mu_{B}$ found in the alkali atoms which are the standard workhorses used in most ultracold atomic experiments. While the dominant interatomic interactions in alkali atom condensates are the isotropic van der Waals interactions which fall off as $1/r^{6}$, the effects of magnetic dipole-dipole interactions, which are anisotropic and fall off as $1/r^{3}$, have been seen in the BECs composed of the more strongly magnetic atoms listed above (note that the strength of the magnetic dipole-dipole interaction goes as $d^{2}$). In particular, magnetostriction of a condensate has been observed \cite{Lahaye2007} as well as {\it d}-wave collapse and explosion \cite{Lahaye2008}.  Of particular relevance to this paper are theoretical studies of dipolar condensates which predict roton excitations \cite{ODell2003,Santos2003,Ronen07}, akin to those observed in Helium II. Although the experimental observation of rotons in dipolar BECs with static dipole-dipole interactions remains illusive, their effects  have been seen in BECs with laser-induced dipole-dipole interactions \cite{Mottl2012}. For a comprehensive review of the field of dipolar BECs up to 2009 we refer the reader to the article \cite{Lahaye2009}.

Despite the enhanced dipole-dipole interactions in $^{52}$Cr relative to alkali atoms, the dipolar interactions in $^{52}$Cr  still play a secondary role compared to the van der Waals interactions unless something is done to suppress the latter. Fortuitously, this has proved possible using the technique of Feshbach resonance \cite{Werner05} which has allowed the creation of a purely dipolar condensate \cite{koch}. This method has also been applied to $^{168}$Er \cite{Aikawa}. Polar molecules are another promising system for observing (electric) dipolar interactions due to their potentially very large electric dipole moments. These systems are now close to quantum degeneracy  \cite{Ni}.

The quantum-coherent nature of BECs means that they support vortices characterised by a core of vanishing density and a 2$\pi$-quantized singularity of the quantum mechanical phase.  Diverse vortex structures have been generated in BECs, from single vortices to vortex rings and vortex pairs, to vortex lattices and turbulent states \cite{Emergent,Fetter}.  While vortices have not yet been observed in dipolar BECs, simulations indicate that vortex rings were formed in the {\it d}-wave collapse experiment \cite{Lahaye2008}.  Theoretical work has also revealed that dipolar interactions modify the vortex properties, introducing density ripples about the core  and an elliptical core \cite{pu,Wilson,abad,Mulkerin2013}, as well as affecting the critical rotation frequency for vortex generation \cite{ODell2007} and the associated dynamical instability \cite{Bijnen2007,Bijnen2009,Kishor2012}. The critical obstacle velocity for vortex nucleation is likewise modified \cite{Ticknor2011} as well as the  structure of vortex lattices  \cite{lattices}.

Here we examine the single vortex solutions in an infinite two-dimensional dipolar Bose gas where the dipoles are aligned by an external field, expanding on some aspects of our recent work \cite{Mulkerin2013}.  We begin by establishing the stability regimes of the homogeneous 2D system.  We then explore the vortex solutions as a function of dipolar strength and polarization angle, and end with an analysis of the dipolar potential generated by the vortex.

\section{Mean-Field Model}

We consider a dilute BEC composed of atoms of mass $m$ with dipole moment $d$.  These atomic dipoles are polarized in a common direction by an external magnetic field, with the polarization axis taken to lie at angle $\alpha$ to the $z$-axis and in the $xz$ plane.  For the purposes of Gross-Pitaevskii mean-field  theory the total interaction between two atoms at ${\bf r}$ and ${\bf r}'$ can be represented by the potential \cite{Lahaye2009,yi01},
\begin{equation}
U(\mathbf{r-r'})=g_{\rm 3D} \delta(\mathbf{r}-\mathbf{r'}) + \frac{C_{\rm dd}}{4\pi}\frac{1-3\cos^{2}\theta}{|\mathbf{r-r'}|^3}.
\label{eqn:atomic_interaction}
\end{equation}
The first term is a pseudo-potential which models the van der Waals interactions through a contact potential with coupling strength $g_{\rm 3D}=4 \pi \hbar^2 a_{\rm s}/m$, where $a_{\rm s}$ is the {\it s}-wave scattering length.  The second term is the bare dipole-dipole interaction, where $\theta$ is the angle between the polarization direction and the inter-particle vector ${\bf r}-{\bf r'}$.  For magnetic dipoles the dipolar coupling strength is $C_{\rm dd}=\mu_0 d^2$, where $\mu_0$ is the permeability of free space.  

Working in the limit of zero temperature, we describe the condensate by the mean-field wavefunction $\Psi({\bf r},t)$, which specifies the atomic density $n({\bf r},t)=|\Psi({\bf r},t)|^2$.  The wavefunction obeys the dipolar Gross-Pitaevskii equation,
\begin{equation}
i \hbar \frac{\partial \Psi}{\partial t}=\left [-\frac{\hbar^2}{2m}\nabla^2+V({\bf r})+g_{\rm 3D}|\Psi|^2+\Phi_{\rm 3D}(\boldsymbol{r},t) - \mu_{\rm 3D}\right ]\Psi,
\label{eqn:dgpe1}
\end{equation}
where $V({\bf r})$ is the external potential acting on the condensate and $\mu_{\rm 3D}$ is the chemical potential.
The van der Waals interactions lead to the purely local term $g_{\rm 3D}|\Psi|^2$ in the above Hamiltonian, while the dipolar interactions are incorporated via the non-local potential $\Phi_{\rm 3D}$ given by
\begin{equation}
\Phi_{\rm 3D}(\boldsymbol{r},t)=\int U_{\rm dd}(\boldsymbol{r-r}')n(\boldsymbol{r'},t)~d\boldsymbol{r}',
\end{equation}
where $U_{\rm dd}$ is the dipolar interaction potential, the second term in Eq. (\ref{eqn:atomic_interaction}).

\begin{figure}[h]
\centering
	\includegraphics[width=0.45\columnwidth,angle=0]{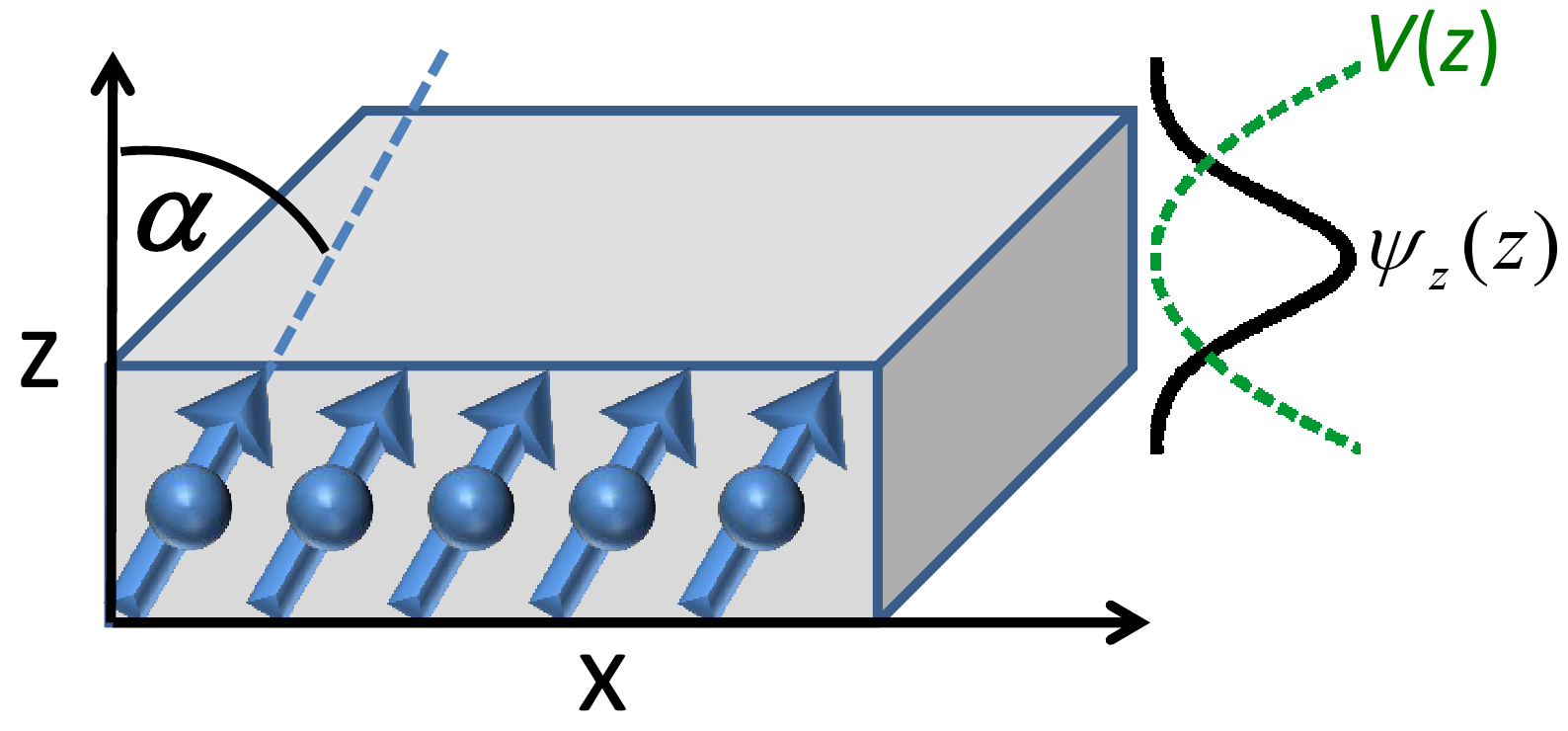}
		\caption{Schematic of our homogeneous 2D dipolar gas. The dipoles are taken to be polarized at angle $\alpha$ to the $z$-axis in the $xz$-plane.}
		\label{fig:schematic}
\end{figure}

The reduction to an effectively 2D BEC, homogeneous in one plane, is performed by considering the external potential to be harmonic trapping in the axial direction only, $V({\bf r})=(1/2)m \omega_z^2 z^2$, where $\omega_z$ is the corresponding trap frequency.  The axial trapping is assumed to be so strong that $\Psi$ becomes frozen into the axial ground harmonic state, a Gaussian of width $l_z=\sqrt{\hbar/m \omega_z}$.  Formally, this ``2D mean-field regime" is reached if $(1+2 \varepsilon_{\rm dd})a_s l_z n \ll 1$, where $n$ is the average 2D density \cite{Parker}.  This scenario of an effectively 2D homogeneous dipolar BEC is illustrated in Fig. \ref{fig:schematic}.  

Within the 2D mean-field regime one can decompose the wavefunction according to $\Psi({\bf r},t)=\psi(\boldsymbol{\rho},t)\psi_z(z)$, where $\psi$ is the 2D wavefunction in the $xy$-plane.  Integrating over $z$ then leads to the effective 2D dipolar GPE \cite{Ticknor2011}, equivalent to Eq. (\ref{eqn:dgpe1}) with the following 3D quantities replaced by their 2D analogs: $\mu_{\rm 3D} \mapsto \mu$, $g_{\rm 3D} \mapsto g=g_{\rm 3D}/\sqrt{2\pi}l_z$, $\Phi_{\rm 3D} \mapsto \Phi$ and $\Psi({\bf r},t) \mapsto \psi(\boldsymbol{\rho},t)$.  The reduced dipolar potential can be expressed as,
\begin{equation}
\Phi(\boldsymbol{\rho},t)=\int U^{\rm 2D}_{\rm dd}(\boldsymbol{\rho-\rho}')n(\boldsymbol{\rho'},t)~d\boldsymbol{\rho}',
\label{eqn:2d_phi}
\end{equation} 
where $U^{\rm 2D}_{\rm dd}$ is the effective 2D dipolar interaction potential.  Rather than evaluating $\Phi$ directly through Eq.~(\ref{eqn:2d_phi}), it is more convenient to evaluate it using the convolution theorem as $\Phi(\boldsymbol{\rho},t)=\mathcal{F}^{-1} \left[\tilde{U}^{\rm 2D}_{\rm dd}({\bf k}) \tilde{n}({\bf k},t)\right]$.  Here the Fourier transform of $U^{\rm 2D}_{\rm dd}$ has been shown elsewhere to be \cite{Ticknor2011,Fischer2006},
\begin{equation}
\tilde{U}^{\rm 2D}_{\rm dd}(\mathbf{q})=\frac{4\pi g_{\rm dd}}{3} \left[F_{\parallel}\left (\mathbf{q} \right ) \sin^2\alpha +F_{\perp}\left (\mathbf{q} \right )\cos^2\alpha\right],
\end{equation}
with $g_{\rm dd}=C_{\rm dd}/3 \sqrt{2 \pi}l_z$, $\mathbf{q}=\mathbf{k}l_z/\sqrt{2}$, $F_{\parallel}({\bf q})=-1+3\sqrt{\pi}\frac{q_{x}^{2}}{q}e^{q^2} \mathrm{erfc}(q)$ and $F_{\perp}({\bf q})=2-3\sqrt{\pi}q e^{q^2} \mathrm{erfc}(q)$.  The axial width of the condensate $l_z$ plays a more important role for 2D dipolar condensates than their non-dipolar counterparts since it specifies how much of the out-of-plane dipolar interaction is experienced by the condensate.

 The strength of the dipolar interactions can be parameterized via the ratio \cite{Lahaye2009} 
 \begin{equation}
 \varepsilon_{\mathrm{dd}}=g_{\mathrm{dd}}/g \ .  
 \end{equation}
We note that $g$ can be tuned between $-\infty$ and $+\infty$ via a Feshbach resonance while $g_{\rm dd}$ can be varied below its natural value, including to negative values, via magnetic field rotation \cite{Lahaye2009}.  Hence it is feasible to consider $-\infty < \varepsilon_{\rm dd} < \infty$ and both negative and positive $g_{\rm dd}$.  

Numerical solutions are obtained by propagating the 2D dipolar GPE in imaginary time using a Runga-Kutta algorithm on a  $256^2$ grid extending over a region $[-50,50]\xi_0 \times [-50,50]\xi_0$, where $\xi_0$ is the healing length (defined in the following section).   At the boundary of the numerical box, the value of the wavefunction is fixed to the uniform value to mimic an infinite system (Dirichlet boundary conditions).

\section{The Homogeneous 2D Dipolar Gas}

In the absence of any potential in the plane the stable ground state solutions have uniform density, which we denote $n_0$.  This generates a uniform dipolar potential given by $\Phi_0 = n_0 g_{\rm dd}(3\cos^2 \alpha -1)$.  The dipolar GPE then reduces to \cite{Ticknor2011},
\begin{equation}
\mu_0=n_0 \left(g+g_{\rm dd} [3\cos^2\alpha-1]\right ),
\end{equation}
where $\mu_0$ is the chemical potential of the homogeneous system.  This relation is similar to the result for a homogeneous 2D non-dipolar BEC $\mu = n_0 g$, but where the dipolar interactions effectively introduce an additional contribution to the contact interaction $g_{\rm dd} [3\cos^2\alpha-1]$.  The angular dependence accounts for the ``tilt" of the dipoles.  Taking the dipoles to be conventional in the sense that $g_{\rm dd}>0$, then for $\alpha=0$ the dipoles lie side-by-side and experience the repulsive part of the dipole-dipole interaction, while for $\alpha=\pi/2$, the dipoles lie head-to-tail in the plane and experience the attractive part of the interaction.  For $g_{\rm dd}<0$, the sign of the dipole-dipole interaction becomes reversed, such that the $\alpha=0$ arrangement of dipoles becomes {\it attractive} and the $\alpha=\pi/2$ arrangement becomes {\it repulsive}.  Note that, at the ``magic angle" $\alpha_0 = \arccos(1/\sqrt{3})\approx 54.7^\circ$, the in-plane dipolar interaction is zero (for either sign of $g_{\rm dd}$).    

A natural length scale for the homogeneous system is the dipolar healing length, $\xi_0=\hbar/\sqrt{m \mu_0}$.  In our analysis, we set the axial length scale of the condensate to be $l_z=0.5 \xi_0$, which satisfies the criteria for the system to be two-dimensional $l_z/\xi_0<1$.

\subsection{The phonon instability}
\label{sec:PI}

Perturbations in the density and velocity of the condensate, with frequency $\omega$ and wavevector $k$, satisfy the Bogoluibov excitation spectrum \cite{Lahaye2009},
\[E(k)=\hbar k \sqrt{\frac{\mu_0}{m} +\frac{\hbar^2 k^2}{4m^2}}. \]
When $\mu_{0}$ is less than zero (i.e. the net local contract interactions are attractive), then in the phonon limit $k \rightarrow 0$, the perturbations possess imaginary energies, signifying the unstable exponential growth of the perturbations over time.  This is the {\it phonon instability}, familiar from the conventional condensate with attractive contact interactions \cite{Pethick2002}.

\begin{figure}[t]
\centering
	\includegraphics[width=0.9\columnwidth,angle=0]{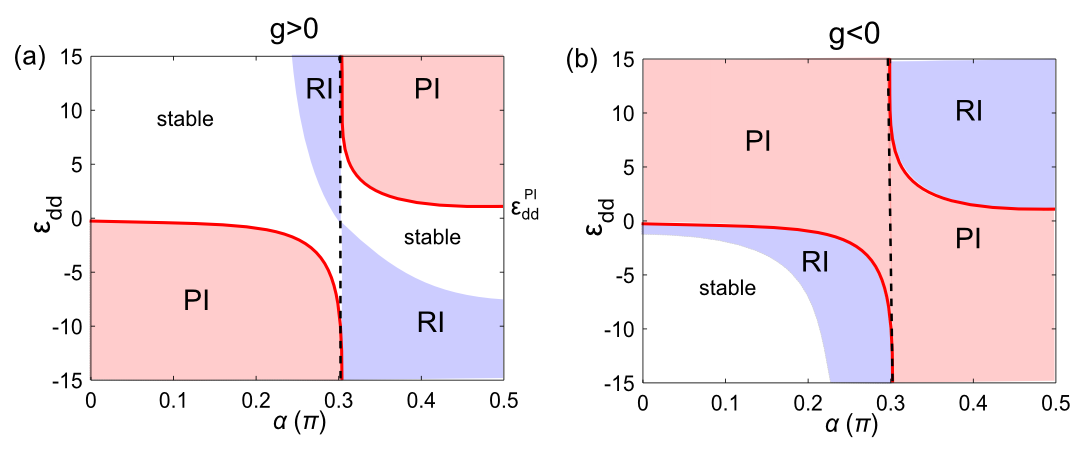}
		\caption{Stability diagram  in $\varepsilon_{\rm dd}-\alpha$ space for the homogeneous 2D dipolar BEC ($\sigma=0.5$) with (a) $g>0$ and (b) $g<0$.  Shown are the regions of stability (white), phonon instability (pink) and roton instability (blue).  The threshold for the phonon instability $\varepsilon_{\rm dd}^{\rm PI}=[1- 3 \cos^2 \alpha ]^{-1}$ is shown by the red lines.  The vertical dashed line indicates the magic angle $\alpha_0$.  Note that the plots only extend up to $\alpha=\pi/2$; beyond this the results are the mirror image of the presented region.   }
		\label{fig:phase_diagram}
\end{figure}

From Eq. ($7$), the instability condition $\mu_0<0$ means that \cite{Santos2003,Fischer2006,Lahaye2009,Bijnen},
\begin{equation}
g+g_{\rm dd} [3 \cos^2 \alpha-1] < 0.
\label{eqn:crit}
\end{equation}
Let us consider the implications for the cases of $g>0$ and $g<0$ separately, with the results summarized in Fig. \ref{fig:phase_diagram}(a) and (b) respectively.  For $g>0$ the system undergoes the phonon instability (PI) when the in-plane dipolar contact interactions $g_{\rm dd} [3\cos^2\alpha-1]$ are attractive (when either $g_{\rm dd}>0$ and $\alpha>\alpha_0$, or $g_{\rm dd}<0$ and $\alpha<\alpha_0$) and of sufficient magnitude to overcome the repulsive vdW contact interactions (characterized by $g$).   Dividing Eq. (\ref{eqn:crit}) by $g$ ($>0$) and introducing $\varepsilon_{\rm dd}=g_{\rm dd}/g$ we obtain the condition for the phonon instability to be $1 > -\varepsilon_{\rm dd}[3\cos^2\alpha-1]$.  We define the threshold value of $\varepsilon_{\rm dd}$ as $\varepsilon_{\rm dd}^{\rm PI}= [1- 3 \cos^2 \alpha ]^{-1}$ (red lines in Fig. \ref{fig:phase_diagram}).  Note the importance of the magic angle $\alpha_0$, at which the instability threshold diverges. It then follows that (for $g>0$) the phonon instability arises for $\varepsilon_{\rm dd}< \varepsilon_{\rm dd}^{\rm PI}$ when $\alpha<\alpha_0$, and for $\varepsilon_{\rm dd}> \varepsilon_{\rm dd}^{\rm PI}$ when $\alpha>\alpha_0$, as depicted in Fig. \ref{fig:phase_diagram}(a) (pink shaded regions).

Meanwhile, for $g<0$ the system will be stable to the phonon instability when in-plane dipolar interactions are repulsive and of greater magnitude than the attractive vdW interactions.  The instability criterion is then $\varepsilon_{\rm dd}> \varepsilon_{\rm dd}^{\rm PI}$ for $\alpha < \alpha_0$ and $\varepsilon_{\rm dd}< \varepsilon_{\rm dd}^{\rm PI}$ for $\alpha > \alpha_0$, as depicted by in Fig. \ref{fig:phase_diagram}(b) (pink shaded regions).

\subsection{The roton instability}
\label{sec:RI}
In addition to the phonon instability, a dipolar BEC also experiences the roton instability.  Dipolar interactions in a BEC can induce a roton dip in the excitation spectrum at finite momenta \cite{ODell2003,Santos2003,Wilson}, reminiscent of rotons in superfluid Helium \cite{Donnelly1991}.     For certain parameters, the roton ``softens" to zero energy, inducing a collapse instability at finite momentum, associated with the runaway alignment of dipoles in the head-to-tail configuration (for $g_{\rm dd}>0)$ or side-by-side (for $g_{\rm dd}<0$).  Close to the RI, ripples in the BEC density arise when the roton mode mixes into the ground state \cite{pu,Wilson}.  

We assess the regimes of roton stability/instability by numerically seeking solutions of the 2D dipolar GPE where the initial condition is a homogeneous wavefunction modulated with sinusoidal perturbations in $x$ and $y$,  $\psi(\boldsymbol{\rho},t)=\sqrt{n_0}[1+0.2\sin(x/\xi_0)+0.2\sin(y/\xi_0)]$.  This perturbation facilitates the development of the roton instability, where it exists. Where stationary solutions are not obtained (and the system is in a regime which does not suffer the phonon instability) it is deemed to possess the roton instability.  The roton instability regimes are depicted in the space of $\varepsilon_{\rm dd}-\alpha$ in Fig. \ref{fig:phase_diagram}(a) and (b) (blue shaded regions).

Again, we interpret the $g>0$ and $g<0$ cases separately.  For $g>0$ the roton instability is induced by the attractive part of the dipolar interaction. For  $\alpha=0$ (dipoles polarized parallel to the $z$-axis) the condensate, confined to a narrow region about $z=0$, cannot probe this attractive part of $U_{\rm dd}$ and no roton instability exists \cite{Fischer2006}.  For increased polarization angle $\alpha$ (while still in the regime $\alpha<\alpha_0$), the attractive part of $U_{\rm dd}$ can be experienced by the condensate and so a regime of roton instability does occur.  This falls in the parameter regime $\alpha \gappeq 0.23$ and $\varepsilon_{\rm dd}>0$ (i.e. $g_{\rm dd}>0$).  Meanwhile, for $\alpha>\alpha_0$, a region of roton instability occurs for negative $\varepsilon_{\rm dd}$, i.e. $g_{\rm dd}=g \varepsilon_{\rm dd}<0$.  Although the dipoles are approaching an head-to-tail configuration, since $g_{\rm dd}<0$ the dipoles {\it repel} in the plane and {\it attract} axially, with the latter attraction inducing the roton instability.

Now we consider the roton instability for $g<0$.  For $\alpha=0$ the presence of attractive contact interactions can support a roton and an associated roton instability \cite{Klawunn2009}.  In contrast to the $g>0$ case, for $g<0$ the roton instability exists for all $\alpha$ (barring the magic angle).  For $\alpha<\alpha_0$ the roton instability occurs in a band of $\varepsilon_{\rm dd}$ in the negative $\varepsilon_{\rm dd}$ half-plane, i.e. $g_{\rm dd}=g \varepsilon_{\rm dd} >0$.  As for $\alpha=0$, the roton instability here is induced by the attractive contact interactions.  As $\varepsilon_{\rm dd}$ is made increasingly negative (corresponding to increasingly positive $g_{\rm dd}$) the roton instability is removed due to the stabilizing effect of the large, repulsive in-plane dipolar interactions.  For $\alpha>\alpha_0$, a region of roton instability occurs for positive $\varepsilon_{\rm dd}$, i.e. negative $g_{\rm dd}$.  As for $g>0$, $\varepsilon_{\rm dd}<0$ case discussed above, the roton instability here is driven by the attractive axial component of the dipolar interactions.

Looking at the overall stability/instability of the system depicted in Fig. \ref{fig:phase_diagram}, it is noteworthy that, for $g>0$, regimes of stability exist for all $\alpha$, while for $g<0$ the only stable regime lies for small $\alpha$ and negative $\varepsilon_{\rm dd}$.

\subsection{Role of $\sigma$}
It follows from the analysis in Section \ref{sec:PI} that the regimes of phonon stability/instability, and the threshold between them, is independent of the axial size of the condensate, $\sigma$.  However, the roton instability is affected by $\sigma$.  For a narrower (wider) condensate, the regimes of roton instability in the phase diagram shrink (expand).  This is because the atoms experience less (more) of the out-of-plane component of $U_{\rm dd}$.  To illustrate the quantitative dependence, take for example $\alpha=0$.  Denoting the threshold for the roton instability as $\varepsilon^{\rm RI}_{\rm dd}$ the solutions are unstable for $\varepsilon^{\rm RI}_{\rm dd}(\sigma)<\varepsilon_{\rm dd}<\varepsilon^{\rm PI}_{\rm dd}$, where $\varepsilon^{\rm PI}_{\rm dd}=-0.5$.  For $\sigma=0.2$, $0.5$ and $1$ then $\varepsilon^{\rm RI}_{\rm dd}\approx -0.7$, $-1.2$ and $-2.7$, respectively.

\section{Vortex Solutions}
\label{sec:vortex_solutions}

We generate singly-charged vortices at the origin of our otherwise homogeneous system by setting the phase $S(\boldsymbol{\rho})=\arctan \left[{\rm Im}(\psi)/{\rm Re}(\psi)\right]$ of the wavefunction  to be $S= \arctan[y/x]$ (corresponding to a singly-charged vortex) during imaginary time propagation of the dipolar GPE.  

In the non-dipolar system, a vortex has a size characterized by the conventional healing length $\xi_0=\hbar/\sqrt{m \mu_0}$, where $mu_0=n_0 g$.  In the dipolar system, the vortex core size is of the order of the dipolar healing length $\xi_0$.  This is the motivation for expressing length in these units.  However, it should be noted that this length scale is a function of the system parameters, and even diverges as $\mu_0 = n_0 (g+g_{\rm dd} [3\cos^2\alpha-1]) \rightarrow 0$.   

\begin{figure}[b]
	\includegraphics[width=1\columnwidth,angle=0]{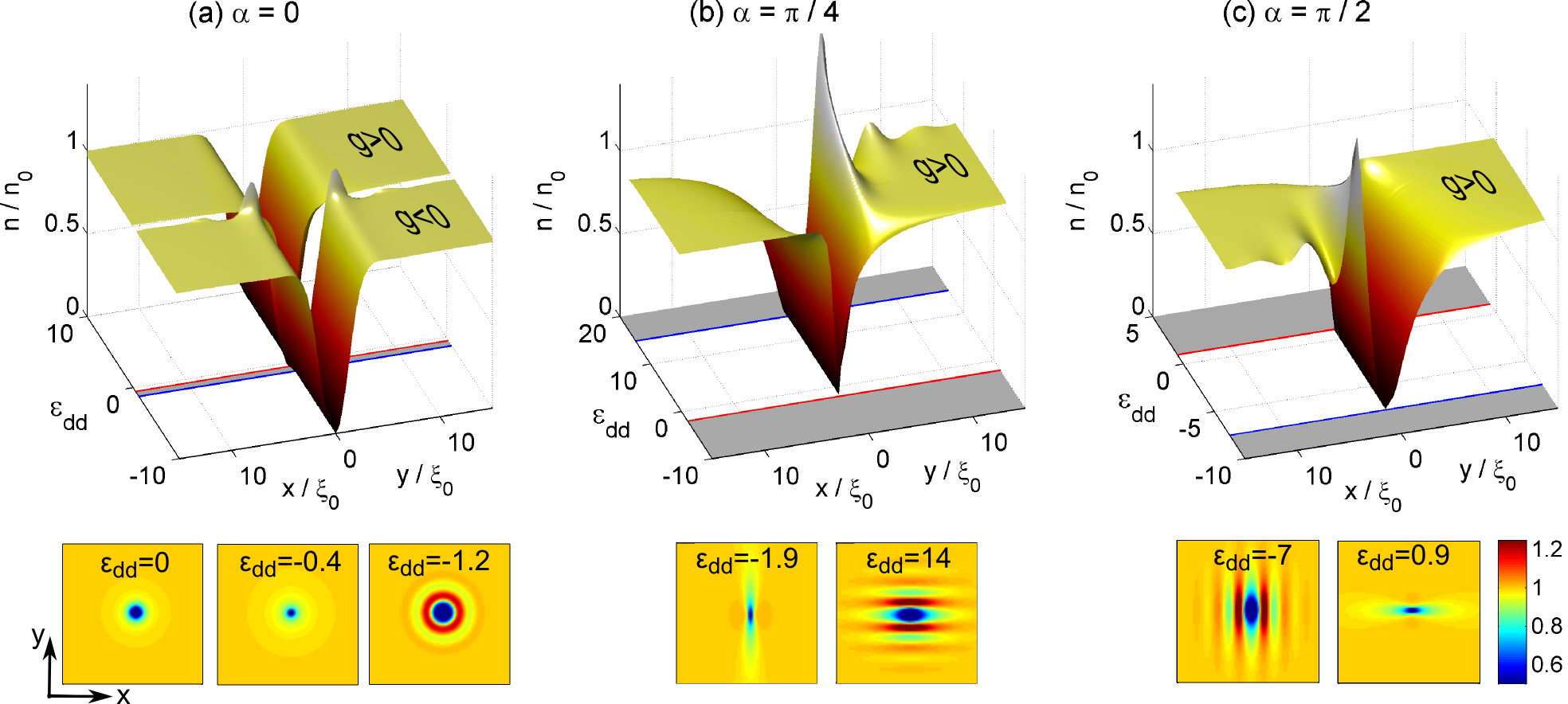}
		\caption{Vortex density profile in the presence of dipolar interactions $\varepsilon_{\rm dd}$ (with $\sigma=0.5$).    The left (right) side of the abcissa  shows the profile along $x$ ($y$).  (a) Dipoles polarized along $z$ ($\alpha=0$).   (b) Dipoles polarized off-axis at $\alpha=\pi/4$.  (c) Dipoles polarized off-axis at $\alpha=\pi/2$.  Grey bands indicate the unstable regimes of $\varepsilon_{\rm dd}$.  Insets: vortex density $n(x,y)$ for example values of $\varepsilon_{\rm dd}$ over an area $(40\xi_0)^2$.
}
		\label{fig:single_vortices}
\end{figure}

We examine the structure of the vortex solutions for three representative polarization angles: (a) $\alpha=0$, (b) $\pi/4$ and (c) $\pi/2$ (each of which fall into distinct regimes of the stability diagram).    Figures~\ref{fig:single_vortices}(a), (b) and (c) plot the vortex solutions for the respective polarization angle, as a function of $\varepsilon_{\rm dd}$.  The left-hand side of the main plots shows the density profile along $x$ (with $y=0$) and the right-hand side shows the density profile along $y$ (with $x=0$).  Note that we find that the introduction of a vortex has no observable effect on the stability of the system, which maintains the stability diagram plotted in Fig. \ref{fig:phase_diagram}, and hence regions with no vortex solutions occur in the plot.

\begin{description}
\item \underline{(a) $\alpha=0$:}    Since the dipoles are perpendicular to the plane, the dipolar potential, and hence density profile, are circularly-symmetric [\ref{fig:phase_diagram}(a)].  For $\varepsilon_{\rm dd}=0$ the vortex has the standard appearance familiar from non-dipolar condensates [see left inset of Fig.\ \ref{fig:phase_diagram}(a)], consisting of a circularly-symmetric core of vanishing density of width $\xi_0$ \cite{Pethick2002}.

For $\varepsilon_{\rm dd} \neq 0$ and for most of the parameter space the vortex solution is essentially identical to that for $\varepsilon_{\rm dd}=0$   (due to our choice of $\xi_0$ as the length scale).  Here the system behaves like a non-dipolar system but with modified contact interactions, and the non-local effects of the dipolar interaction are insignificant.  However, there exist two limiting regimes where the vortex solutions deviate significantly.  Firstly, as one approaches $\varepsilon_{\rm dd}=-0.5$ from above, the vortex core becomes increasingly narrow (in units of $\xi_{0}$) and takes on a modified profile [see middle inset of Fig.\ \ref{fig:phase_diagram}(a)]. This is associated with the cancellation of explicit contact interactions in the system---the vdW interactions cancel the contact contribution from the dipolar interactions.  Indeed, the total contact interaction disappears altogether at the PI threshold, $\varepsilon_{\rm dd}=-0.5$).  Secondly, as the RI is approached from below, circularly-symmetric density ripples appear around the vortex [see middle inset of Fig. \ref{fig:phase_diagram}(c)]. The ripples decay with distance from the core and have an amplitude of up to $\sim 20\% n_0$.   These ripples are associated with the roton mode mixing into the ground state \cite{pu,Wilson}, and are physically associated with the energetic favourability of dipoles aligning head-to-tail, but where the van der Waals interactions and axial trapping prevent collapse.   

Such ripples have been predicted to appear around vortices in trapped 3D dipolar condensates  \cite{pu,Wilson,abad}, and around more general localized density perturbations \cite{Ticknor2011}.  The ripple wavelength, which reflects the roton wavelength, is approximately $4 \xi_0$.

\item \underline{(b) $\alpha=\pi/4$:} The dipoles are now polarized off-axis, i.e.\ tipped in the direction of the $x$-axis.  As a result, the circular-symmetry of the dipolar potential and density is lost, as evident from the asymmetry of the family of solutions in Fig.~\ref{fig:single_vortices}(b).  For $g_{\rm dd}>0$ ($<0$) the dipoles attract (repel) along $x$ and repel (attract) along $y$.  The local dipolar potential is still net positive throughout (since $\alpha < \alpha_0$). The vortex core is elongated along $x$ for $\varepsilon_{\rm dd}>0$, and reverses for $\varepsilon_{\rm dd}<0$.   Close to the RI, density ripples form about the vortex (of large amplitude up to $\sim 40\% n_{0}$) with wavefronts parallel to the $x$-axis.  This alignment is due to the preferred head-to-tail alignment of dipoles which are tipped along $x$, and related to the {\em anisotropic} mixing of the roton into the ground state \cite{Ticknor2011}.

\item \underline{(c) $\alpha=\pi/2$:}  For $g_{\rm dd}>0$ ($<0$) the dipoles are net attractive (repulsive) along $x$ and repulsive (attractive) along $y$.  The anisotropy of the vortex core and ripples is reversed compared to $\alpha=\pi/4$.
\end{description}

\section{Dipolar Potential of a Vortex}

\subsection{Dipoles Perpendicular to the Plane ($\alpha$=0)}

A region of depleted density due to the presence of a vortex in an otherwise uniform system may be alternatively viewed as a lump of ``anti-dipoles" (whose directions have been reversed) sitting in empty space \cite{klawunn,ODell2007}.  From this point of view a vortex  treated as an object in its own right will generate its own mesoscopic dipolar potential. We now seek to estimate this dipolar potential analytically and compare our estimate against exact numerical calculations. 

\begin{figure}[h]
\centering
	\includegraphics[width=0.7\columnwidth]{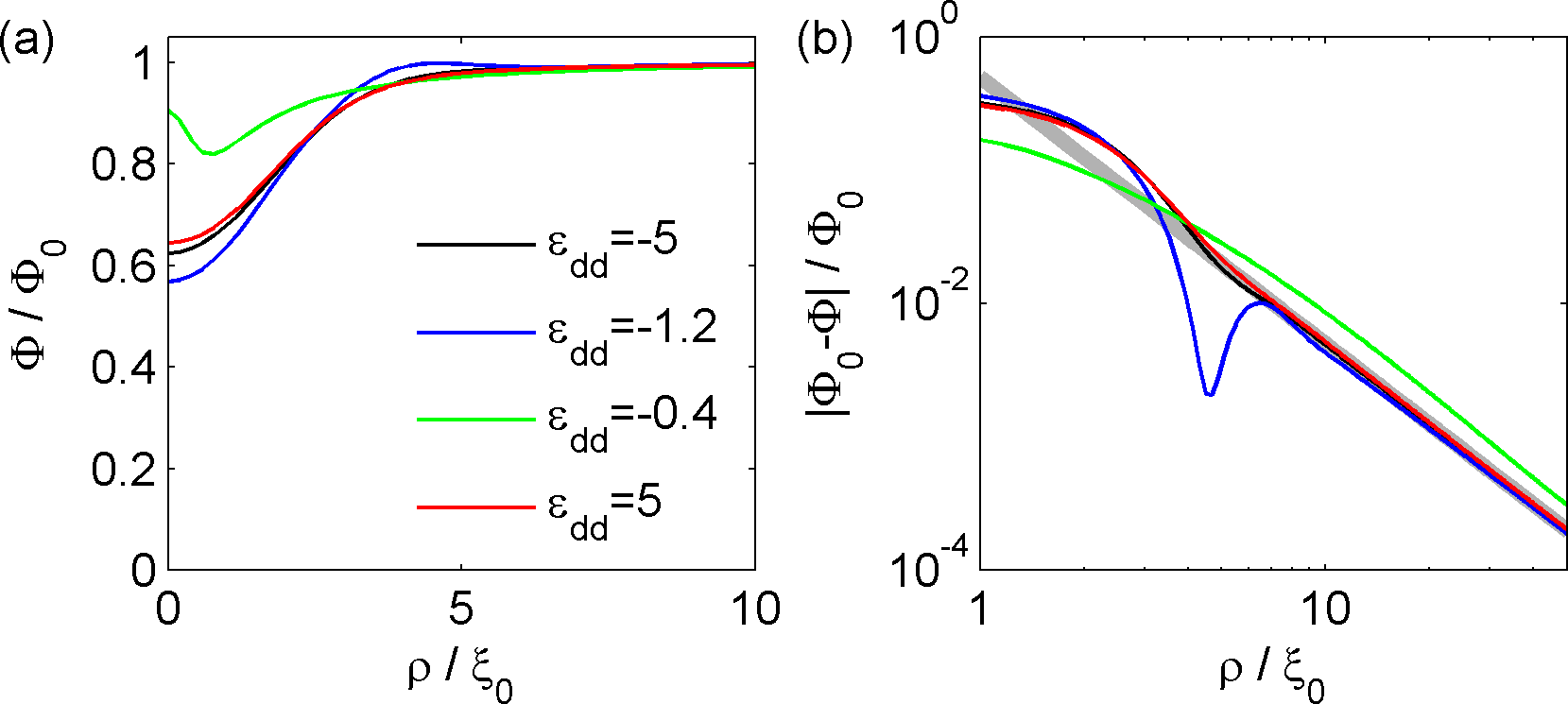}
	\caption{Vortex dipolar potential $\Phi(\rho)$, rescaled by the homogeneous background value $\Phi_0$, for $\alpha=0$ and various values of $\varepsilon_{\rm dd}$ (and $\sigma=0.5$).  The derived $1/\rho^2$ result of Eq.~(\ref{eq:analytic_phidd}) is shown (thick grey line).  The results for $\varepsilon_{\rm dd}=-5$ and $\varepsilon_{\rm dd}=5$ closely match since the vortex profiles are practically identical to each other (approximating the non-dipolar vortex).  The behaviour differs for $\varepsilon_{\rm dd}=-0.4$ and $\varepsilon_{\rm dd}=-1.2$, close to the phonon and roton instabilities respectively, where the vortex develops a modified structure. In (b) we see that, for all cases, the dipolar potential decays as $1/\rho^2$ at long-range.  }
	\label{fig:vortex_dipolar_potential}
\end{figure}

Consider first the simplified case of dipoles polarized perpendicular to the plane, $\alpha=0$. In Fig.~\ref{fig:vortex_dipolar_potential}(a) we show the dipolar potential $\Phi(\rho)$ in the vicinity of the vortex core for various values of $\varepsilon_{\rm dd}$.  These values range from large negative $\varepsilon_{\rm dd}$ ($\varepsilon_{\rm dd} =-5$, black line) to  large positive $\varepsilon_{\rm dd}$ ($\varepsilon_{\rm dd} =5$, red line).  In between we include a case close to the roton instability where the vortex possesses large ripples ($\varepsilon_{\rm dd}=-1.2$, blue line) and close to the phonon instability, where the vortex core becomes narrow ($\varepsilon_{\rm dd}=-0.4$, green line).

Far away from the core $\Phi$ approaches $\Phi_0 = n_0 g_{\rm dd}(3\cos^2 \alpha -1)$, the homogeneous result.  For $\rho \lappeq 5\xi_0$, $\Phi$ is dominated by the core structure and ripples (where present).  Figure ~\ref{fig:vortex_dipolar_potential}(b) plots, on a log-log scale, the decay of $\Phi$ towards the homogeneous value.  Far away from the core, $\Phi$ decays as $1/\rho^2$ (grey line).

The long-range behaviour of the dipolar potential can be obtained analytically from  the vortex ansatz \cite{Fetter_vortex,ODell2007},
\begin{eqnarray}
n(\rho)=n_0\left(1- \frac{1}{1+\rho'^2}\right),
\label{eqn:ansatz}
\end{eqnarray}
where $\rho'=\rho/\xi_0$.  This ansatz gives both the correct $\rho'^{2}$ scaling of the density as $\rho' \rightarrow 0$ and the correct $1/\rho'^2$ decay to the homogeneous density as $\rho' \rightarrow \infty$ \cite{Pethick2002}.  
We evaluate the dipolar potential via the convolution result $\Phi(\boldsymbol{\rho},t)=\mathcal{F}^{-1} \left[\tilde{U}^{\rm 2D}_{\rm dd}({\bf k}) \tilde{n}({\bf k},t)\right]$ and by taking advantage of the cylindrical symmetry afforded by the $\alpha=0$ scenario to employ Hankel transformations.  The Hankel transform of Eq. (\ref{eqn:ansatz}) is,
\[\tilde{n}(\mathbf{k})=n_0 \left(\frac{\delta (k)}{k}+\frac{K_0( k / \sqrt{2})}{2}\right),\]
where $K_0(k)$ is a modified Bessel function of the second kind. We expand the dipolar interaction $\tilde{U}^{\rm 2D}_{\rm dd}(\mathbf{k})$ in terms of the condensate width parameter $\sigma$ up to first-order giving,
\[\tilde{U}^{\rm 2D}_{\rm dd}(\mathbf{k}) = g_{\rm dd}\left(2-\sqrt{\frac{9 \pi }{2}} k \sigma \right) +\mathcal{O}\left(\sigma ^2\right).\]
Then, up to first order in $\sigma$ and third order in $1/\rho'$, the dipolar potential generated by the vortex ansatz is,
\begin{alignat}{1}
\frac{\Phi(\mathbf{\rho})}{\Phi_0}= \left(1 -\frac{1}{\rho'^{2}}\right)+\left(\frac{A \ln \rho' + B}{\rho'^{3}}\right) \sigma,\label{eq:analytic_phidd}
\end{alignat}
with constants $A=-\sqrt{9\pi/8} \approx -1.88$ and $B=(\ln 2-1)A\approx 0.577$.   Recall $\Phi_0$ is the dipolar potential at infinity.

To analyse this result it is useful to consider $\Phi$ as the sum of a local term $\Phi_{\rm L}$, proportional to density and given by,
\begin{eqnarray}
\Phi_{\rm L}(\boldsymbol{\rho})=n(\boldsymbol{\rho}) g_{\rm dd}(3 \cos^2 \alpha -1),
\end{eqnarray}
and non-local term $\Phi_{\rm NL}$.  The non-local term can be viewed as arising from a fictitious electrostatic potential associated with the dipoles \cite{ODell2004,Parker} and has a complicated form given elsewhere \cite{Bao2010}.  It depends on {\em variations} of density, becoming zero for a homogeneous system.

The first bracketed term in Eq.~(\ref{eq:analytic_phidd}) corresponds to the {\em local} contribution to the vortex dipolar potential.  This is evident since, at long-range ($\rho'\gg1$), the vortex ansatz becomes $n(\rho)=n_0(1-1/\rho'^2)$, i.e.\ it decays as $1/\rho'^2$ to the background density.  Note that the {\it s}-wave interactions generate a local potential with identical form (proportional to density) \cite{Wu1994}.   The second bracketed term in Eq.~(\ref{eq:analytic_phidd}) describes the {\em non-local} contribution, i.e.\ the contribution to the potential arising from the long-range contributions from all anti-dipoles in the vortex.  This vanishes in the true 2D limit $\sigma=0$ since the volume of anti-dipoles in the vortex core vanishes.  This illustrates the point that the dipolar potential generated by a vortex is not a topological quantity like the potential associated with the hydrodynamic flow around the vortex (which leads to the logarithmic interaction between two vortices), but instead depends on the number of dipoles excluded from the region by the presence of the vortex. Also, the fact that the non-local vortex potential scales dominantly as $\ln \rho'/\rho'^3$, and not $1/\rho'^3$, informs us that the vortex does not strictly behave as a point-like collection of dipoles at long-range.  This is due to the slow power-law recovery of the vortex density to $n_0$. This point is reinforced by noting that for sharply-truncated density profiles, e.g.\ a circular ``hole", and exponentially-decaying density profiles, e.g.\ $1-\exp(-\rho'^2)$ or $ \tanh^2 (\rho')$, the dipolar potential \emph{does} decay as $1/\rho'^3$ at long-range.  Due to the fast decay of the density, the inhomogeneity then really does resemble a giant dipole at long-range.  This further confirms the subtle but critical role of the slow power-law decay of the vortex density. Nevertheless, whatever the subtleties of the non-local part of $\Phi$, its long range behaviour is dominated by the local $1/\rho'^2$ scaling.  This is confirmed by our numerical solutions, whose $\Phi$ decays to the background value as $1/\rho'^2$ at long-range, as shown in Fig. \ref{fig:vortex_dipolar_potential}(b).

\subsection{Dipoles with a Projection in the Plane ($\alpha \neq 0$)}

When the polarization direction is tilted away from the vertical, the dipolar interaction becomes anisotropic in the plane, resulting in the anisotropic vortex cores discussed in Section \ref{sec:vortex_solutions} above.  The dipolar potential $\Phi$ will then be anisotropic and  have three contributions: an isotropic contribution from the homogeneous background density, a local contribution from the anisotropic density profile of the vortex, and a non-local anisotropic contribution arising from the anti-dipoles associated with the rest of the vortex.  

\begin{figure}[h]
\centering
	\includegraphics[width=1\columnwidth]{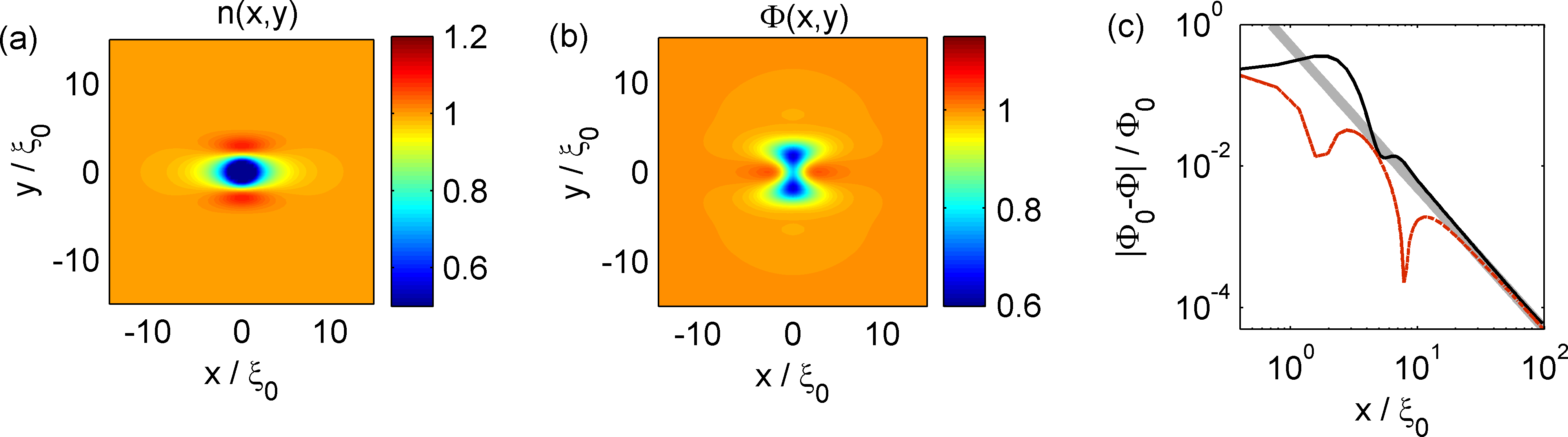}
\caption{Vortex with dipoles polarized off-axis ($\alpha=\pi/4$, $\varepsilon_{\rm dd}=5$ and $\sigma=0.5$).  (a) The density profile $n(x,y)$ is anisotropic, with density ripples polarized along $x$. (b) The dipolar potential $\Phi(x,y)$, rescaled by the homogeneous value $\Phi_0$, is modified from the background value due to the presence of the vortex.  In particular, the collection of anti-dipoles which can be associated with the vortex generates an anisotropic modulation of the dipolar potential which mimics the dipolar potential itself $\sim 1-3\cos^2 \theta$.  (c) The decay of $\Phi$ with distance along $x=0$ (black line) and $y=0$ (red line), on a logarithmic scale.  At short and intermediate scales, the decay varies due to the density variations associated with the vortex, while at long-range the decay approaches the $1/\rho'^2$ form.   The line $1/\rho'^2$ (grey line) is shown for reference. }
	\label{fig:vortex_dipolar_potential2}
\end{figure}

We analyse the ensuing dipolar potential through the exemplar of $\alpha=\pi/4$ and $\varepsilon_{\rm  dd}=5$.  The  density profile of this vortex [Fig. \ref{fig:vortex_dipolar_potential2}(a)] shows a core elongated in $x$ with adjacent density ripples aligned in $x$.  The dipolar potential of the vortex [Fig. \ref{fig:vortex_dipolar_potential2}(b)] is indeed anisotropic, with a raised potential along $x$ and a reduced potential along $y$ (relative to the background dipolar potential $\Phi_0$).  Indeed, the angular dependence of $\Phi$, relative to $\Phi_0$, resembles that of the dipole-dipole interaction itself $\sim 1-3\cos^2 \theta$.  Thus, at least in its angular dependence, the vortex shares qualitatively the characteristics of a mesoscopic dipole.

Away from the core, $\Phi$ decays towards the homogeneous value $\Phi_0= \mu_0 / 2$.  At long-range $\rho' \gg 1$, we again see the decay of $\Phi$ developing a $1/\rho'^2$ behaviour [Fig. \ref{fig:vortex_dipolar_potential2}(c)], consistent with the local contribution to $\Phi$ from the vortex core.  The same behaviour is achieved along the $x$ and $y$ directions; up to mid-scale distances $\rho' \lappeq 10$, $\Phi$ is dominated by the core structure, while at longer ranges, a common $1/\rho'^2$ behaviour is obtained.  

\section{Conclusions}

In this paper we have explored the stability and vortex solutions of the infinite two-dimensional dipolar Bose-Einstein condensate as a function of the strength and polarization angle of the dipoles.  We mapped out the parameter regimes where a homogeneous density distribution in the plane is unstable towards either long wavelength phonon fluctuations or finite wavelength roton fluctuations. The instability is always towards configurations where the dipoles are arranged head-to-tail in order to benefit from the attractive portion of the dipolar interaction. The long-range and anisotropic nature of the dipolar interaction 
can lead to significant modifications to the structure of a single vortex, particularly in and around the core, including the formation of ripples in the core periphery and, for dipoles polarized off-axis, an anisotropic density profile.   Furthermore, the inhomogeneous density of dipoles associated with the vortex generates a  potential which decays slowly from the core and, for dipoles polarized off-axis, varies anisotropically in space.  At long-range, this potential has $1/\rho^2$ and $\ln(\rho)/\rho^3$ contributions, and an angular dependence $\sim 1-3 \cos^2 \theta$.    The long-range dipolar potential emanating from a single vortex in turn modifies the vortex-vortex interaction \cite{Mulkerin2013} and is expected to support new regimes of vortex dynamics and collective structures in many-vortex systems.

\bigskip
DHJO acknowledges support from NSERC (Canada).

\end{document}